\documentclass[]{spie}  

\usepackage{amsmath,amsfonts,amssymb}
\usepackage{graphicx}
\PassOptionsToPackage{hyphens}{url}
\usepackage[colorlinks=true, allcolors=blue]{hyperref}
\usepackage{here}
\usepackage{graphicx}
\usepackage{subcaption}
\usepackage{caption}
\usepackage{siunitx}

\title{Commissioning the CMB polarization telescope GroundBIRD with the full set of detectors}
\author[a]{Miku~Tsujii}
\author[b,c]{Jochem~J.A.~Baselmans}
\author[d]{Jihoon~Choi}
\author[b]{Antonio~H.M.~Coppens}
\author[e]{Alessandro~Fasano}
\author[e]{Ricardo~Tanausú~Génova-Santos}
\author[a]{Makoto~Hattori}
\author[f,g]{Masashi~Hazumi}
\author[h]{Shunsuke~Honda}
\author[i]{Takuji~Ikemitsu}
\author[j,k]{Hidesato~Ishida}
\author[g]{Hikaru~Ishitsuka}
\author[l]{Hoyong~Jeong}
\author[l]{Yonggil~Jo}
\author[b]{Kenichi~Karatsu}
\author[i]{Keisuke~Kataoka}
\author[m]{Kenji~Kiuchi}
\author[i]{Junta~Komine}
\author[n]{Ryo~Koyano}
\author[o]{Hiroki~Kutsuma}
\author[l]{Kyungmin~Lee}
\author[p]{Satoru~Mima}
\author[q]{Makoto~Nagai}
\author[f]{Taketo~Nagasaki}
\author[n]{Masato~Naruse}
\author[r]{Shugo~Oguri}
\author[j,k]{Chiko~Otani}
\author[s]{Michael~W.~Peel}
\author[e]{Rafael~Rebolo}
\author[e]{José~Alberto~Rubiño-Martín}
\author[r]{Yutaro~Sekimoto}
\author[i]{Yoshinori~Sueno}
\author[i]{Junya~Suzuki}
\author[n]{Tohru~Taino}
\author[i]{Osamu~Tajima}
\author[a]{Tomonaga~Tanaka}
\author[b]{David~J.~Thoen}
\author[m]{Nozomu~Tomita}
\author[j,k]{Yuta~Tsuji}
\author[f,g]{Tomohisa~Uchida}
\author[l]{Eunil~Won}
\author[f,g]{Mitsuhiro~Yoshida}

\affil[a]{Astronomical Institute, Tohoku University, Sendai, Miyagi 980-8578, Japan}
\affil[b]{SRON – Netherlands Institute for Space Research, Niels Bohrweg 4, 2333 CA - Leiden, The Netherlands}
\affil[c]{Department of Microelectronics, Faculty of Electrical Engineering, Mathematics and Computer Science (EEMCS), Delft University of Technology, Mekelweg 4, 2628 CD Delft, The Netherlands}
\affil[d]{Korea Astronomy and Space Science Institute, 776 Daedeok-daero, Yuseong-gu, Daejeon 34055, Republic of Korea}
\affil[e]{Instituto de Astrofísica de Canarias, E-38205 La Laguna, Tenerife, Spain}
\affil[f]{The High Energy Accelerator Research Organization (KEK), 1-1 Oho, Tsukuba, Ibaraki 305-0801, Japan}
\affil[g]{The Graduate University for Advanced Studies (SOKENDAI), Shonan Village, Hayama, Kanagawa 240-0193, Japan}
\affil[h]{University of Tsukuba, 1-1-1 Tennodai, Tsukuba, Ibaraki, 305-8571, Japan}
\affil[i]{Kyoto University, Kitashirakawa-Oiwakecho, Sakyo-ku, Kyoto 606-8502, Japan}
\affil[j]{Tohoku University, 2-1-1 Katahira, Aoba-ku, Sendai, Miyagi 980-8577, Japan}
\affil[k]{RIKEN, 519-1399 Aramaki-Aoba, Aoba-ku, Sendai, Miyagi 980-0845, Japan}
\affil[l]{Korea University, 145 Anam-ro, Seongbuk-gu, Seoul, 02841, Republic of Korea}
\affil[m]{The University of Tokyo, 7-3-1 Hongo, Bunkyo-ku, Tokyo 113-8654, Japan}
\affil[n]{Saitama University, 255 Shimo-Okubo, Sakura-ku, Saitama 338-8570, Japan}
\affil[o]{Department of Applied Physics, Tohoku University, Sendai 980-8577, Japan}
\affil[p]{Superconductive ICT Device Laboratory, Kobe Frontier Research Center, Advanced ICT Research Institute, National Institute of Information and Communications Technology, 588-2 Iwaoka, Nishi-ku, Kobe, Hyogo, 651-2492, Japan}
\affil[q]{National Astronomical Observatory of Japan, 2-21-1 Osawa, Mitaka, Tokyo 181-8588, Japan}

\affil[r]{Japan Aerospace Exploration Agency (JAXA), 3-1-1 Yoshinodai, Chuo-ku, Sagamihara-shi, Kanagawa 252-5210, Japan}
\affil[s]{Imperial College London, South Kensington Campus, London SW7 2AZ, UK}

\authorinfo{Further author information: (Send correspondence to Miku Tsujii)\\Miku Tsujii.: E-mail: miku.tsujii.p4@dc.tohoku.ac.jp\\  Makoto Hattori: E-mail: hattori@astr.tohoku.ac.jp}
\begin{document} 
\maketitle

\begin{abstract}
GroundBIRD is a ground-based cosmic microwave background (CMB) experiment for observing the polarization pattern imprinted on large angular scales ($\ell > 6$ ) from the Teide Observatory in Tenerife, Spain. Our primary scientific objective is a precise measurement of the optical depth $\tau$ ($\sigma(\tau) \sim 0.01$) to the reionization epoch of the Universe to cross-check systematic effects in the measurements made by previous experiments. GroundBIRD observes a wide sky area in the Northern Hemisphere ($\sim 40\%$ of the full sky) while continuously rotating the telescope at a high speed of up to 20 rotations per minute (rpm) to overcome the fluctuations of atmospheric radiation.  
We have adopted the NbTiN/Al hybrid microwave kinetic inductance detectors (MKIDs) as focal plane detectors. 
We observe two frequency bands centered at 145 GHz and 220 GHz. The 145 GHz band picks up the peak frequency of the CMB spectrum.  The 220 GHz band helps accurate removal of the contamination of thermal emission from the Galactic interstellar dust.
The MKID arrays (138 MKIDs for 145GHz and 23 MKIDs for 220GHz) were designed and optimized so as to minimize the contamination of the two-level-system noise and maximize the sensitivity. 
The MKID arrays were successfully installed in May 2023 after the performance verification tests were performed at a laboratory.
GroundBIRD has been upgraded to use the full MKID arrays, and scientific observations are now underway.
The telescope is automated, so that all observations are performed remotely.
Initial validations, including polarization response tests and observations of Jupiter and the moon, have been completed successfully. We are now running scientific observations.
\end{abstract}

\section{INTRODUCTION}
\begin{figure}[hbp]
\centering
\begin{subfigure}[b]{0.45\linewidth}
    \centering
    \includegraphics[height=5.cm]{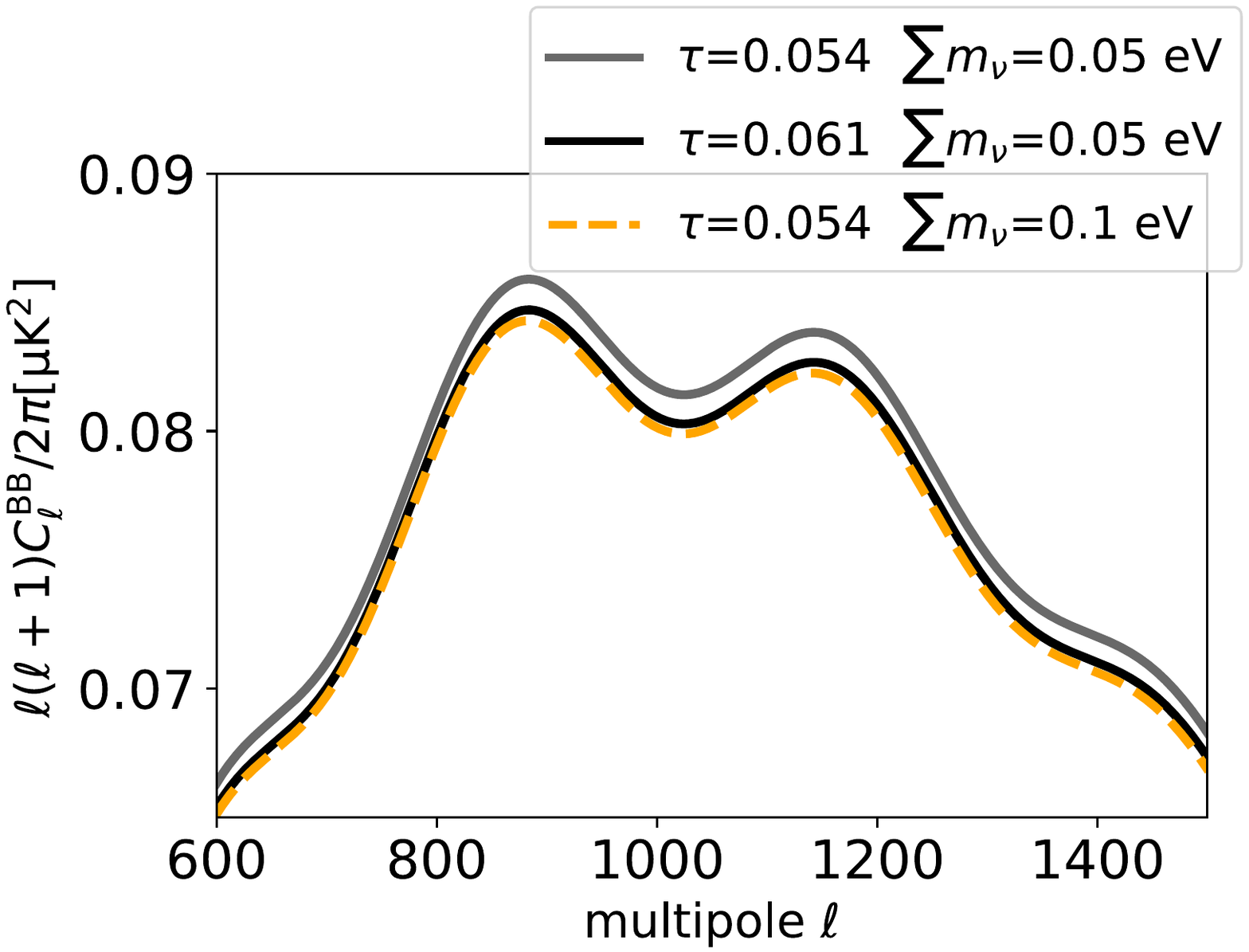}
    \caption{}
    \label{fig:tau_nu_clBB}
\end{subfigure}
\hspace{0.00cm} 
\begin{subfigure}[b]{0.45\linewidth}
    \centering
    \includegraphics[height=5.2cm]{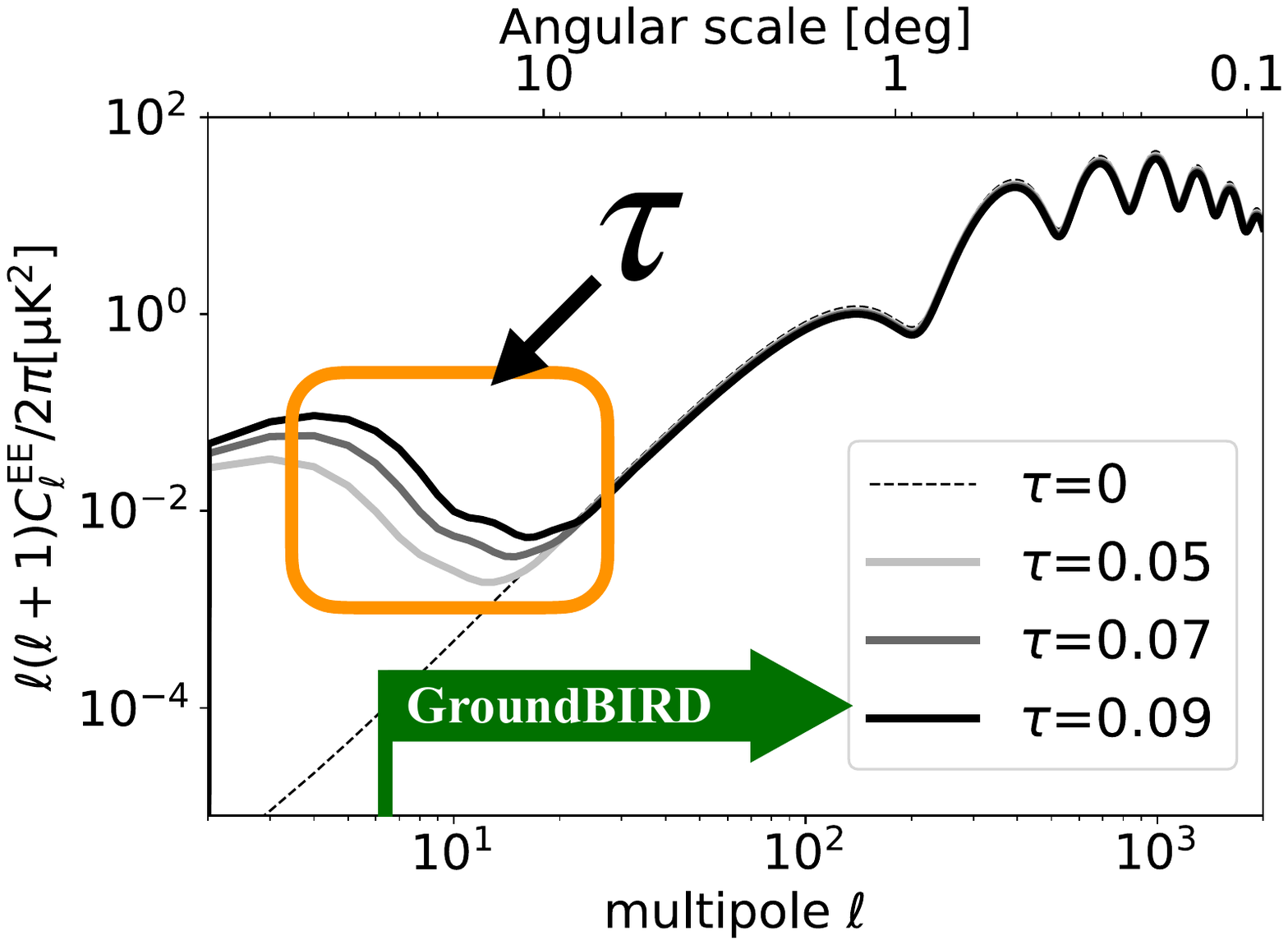}
    \caption{}
    \label{fig:tau_nu_clEE}
\end{subfigure}
\caption[example] 
{\label{fig:tau_nu_clEE&BB}
(a) The polarization power spectrum of the CMB B-mode, calculated using CAMB \cite{lewis2000efficient}. The $\tau$ values used are from Planck Collaboration 2020 \cite{collaboration2020planck} and $1\sigma$ higher. Other cosmological parameters are set as follows: $H_0$ = 67.0, $\Omega_b h^2$ = 0.022, $\Omega_c h^2$ = 0.12, $\Omega_k$ = 0.0, $A_s$ = $2 \times 10^{-9}$, $n_s$  = 0.96, $n_{\text{run}}$ = 0.0, $r$ = 0.0. Even if the total neutrino mass changes to 0.1 eV from 0.05 eV, it is difficult to distinguish this change from the variation in $\tau$.
(b) The polarization power spectrum of the CMB E-mode, calculated using CAMB. The orange range indicates the reionization bumps, where variations occur based on the optical depth $\tau$. GroundBIRD is sensitive to the polarization of CMB at large angular scales ($6 < \ell$).}
\end{figure}

Since the discovery of the cosmic microwave background (CMB)\cite{1965PenziasWilsonCMB,dicke1965cosmic}, the CMB has played a central role in enriching our understanding of the Universe. 
The current prime target of the CMB observation is the B-mode polarization signal imprinted by the primordial gravitational wave generated by the quantum fluctuation of the space-time during the inflation period\cite{kamionkowski1997probe, zaldarriaga1997all, starobinskii1979spectrum, starobinsky1980new, sato1981first, guth1982fluctuations}.
In addition, it has been recognized that precision measurements of the B-mode polarization power spectrum caused by the gravitational lensing effect due to the large-scale structures in the Universe provides a unique opportunity for measuring the total mass of neutrinos\cite{allison2015towards}.  
In Fig.~\ref{fig:tau_nu_clBB}, we show how sensitively the amplitude of the CMB B-mode polarization power spectrum due to the lensing effect depends on the total mass of the neutrinos. 
Figure~\ref{fig:tau_nu_clBB} also shows that the precision measurement of the reionization optical depth, $\tau$, is mandatory to extract the total mass of neutrinos from the B-mode polarization power spectrum. 

There is a systematic difference between the reported values from WMAP ($\tau = 0.089 \pm 0.014$\cite{bennett2013nine})
and from Planck ($\tau = 0.054 \pm 0.007$\cite{collaboration2020planck}) as the precedent. 
The reionization optical depth uncertainties of around 10 $\%$ are relatively high compared with other cosmological parameters, which are measured with subpercent precision\cite{collaboration2020planck}. 
It is mandatory to perform measurements of the reionization optical depth with higher precision using an independent experiment from WMAP and Planck.
However, it is difficult to measure the reionization optical depth from ground-based experiments. 
As shown in Fig.~\ref{fig:tau_nu_clEE}, 
the measurement of the reionization optical depth is performed by using the sensitive dependence of signatures appearing at large angular scales of the power spectrum of the CMB E-mode polarization.
This signature is called the reionization bump
because it is generated by the Thomson scattering of
the CMB photons at the reionization epoch.
The angular scale of the reionization bump corresponds to the 
apparent size of the horizon at the reionization epoch and is about a few tens of degrees.
Due to atmospheric emission fluctuations, the optical loading on the detectors, primarily caused by atmospheric emission, changes significantly while measuring the large angular scale signal.
This makes it difficult to perform precision measurements of the power spectrum of the reionization bump from the ground. 
Therefore, no experiments have reported the reionization optical depth since Planck. 

   \begin{figure} [b]
   \begin{center}
   \begin{tabular}{c} 
   \includegraphics[height=7.5cm]{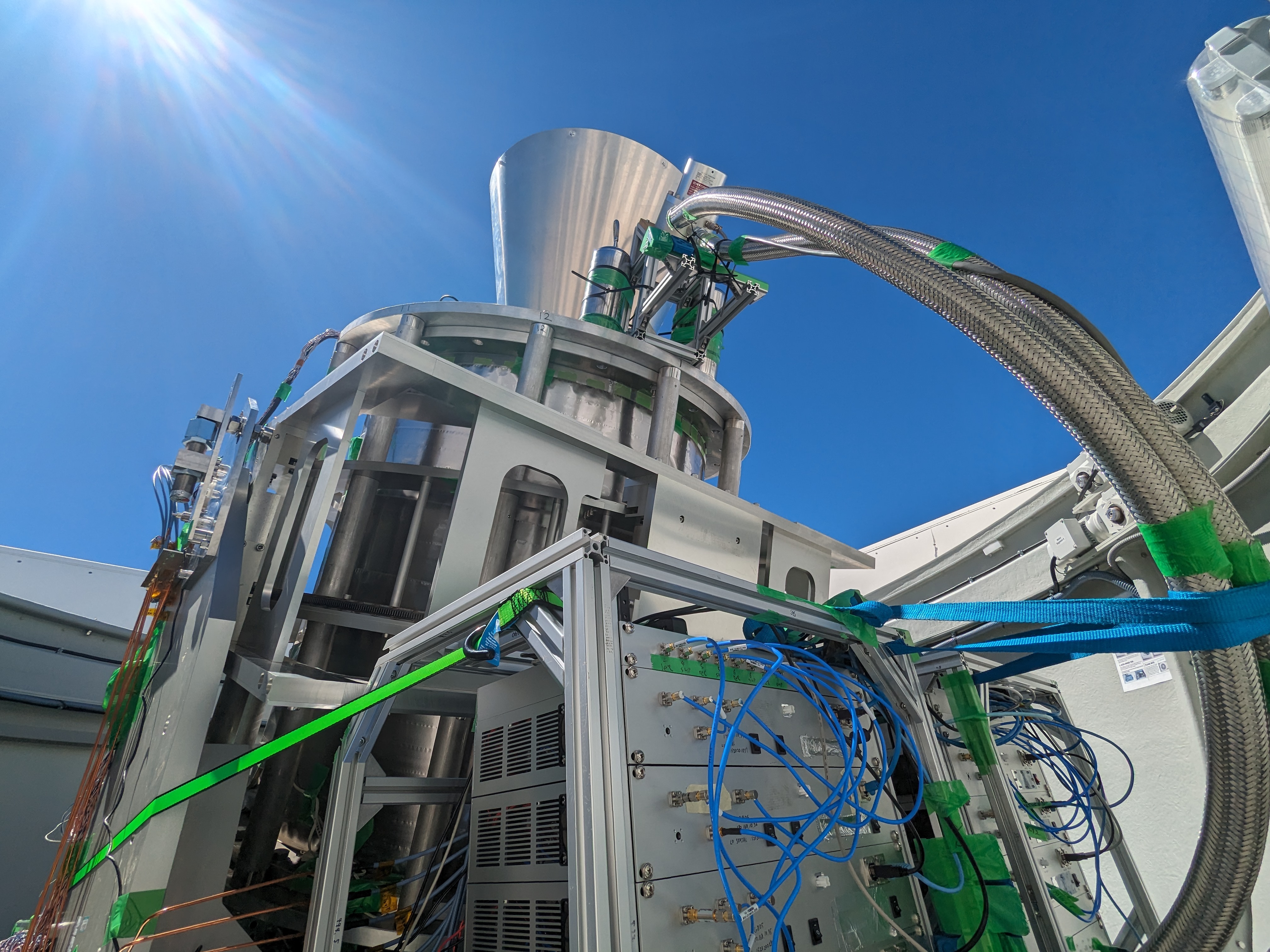}
   \end{tabular}
   \end{center}
   \caption[example]
   {\label{fig:GroundBIRD}
   A photo of GroundBIRD from inside the dome. We see the pulse tube pipe connected at the head next to the optical window and the baffle. The DAQs on the bottom rotate solidly with the telescope.}
   \end{figure} 
GroundBIRD is a ground-based CMB polarization experiment, mounted at the Teide Observatory; $\ang{28;18}$ N and $\ang{16;30}$ W, 2400 m above mean sea level (Fig.~\ref{fig:GroundBIRD}).
To capture the large-scale polarized CMB signal while minimizing the impact of atmospheric 1/f noise, the GroundBIRD telescope employs a high-speed rotation of the telescope around the vertical axis, with the telescope tilted 20 degrees from the zenith, achieving velocities of up to 20 rotations per minute (rpm). This innovative method allows for the observation of over $40\%$ of the celestial sphere.
To realize diffraction-limited spatial resolution while rotating the telescope at high speed, microwave kinetic inductance detectors (MKIDs) are adopted as the focal plane detectors, due to their fast responsivity and high sensitivity. Two frequency bands detectors centered at $\SI{145}{GHz}$ and $\SI{220}{GHz}$ are installed. 
The $\SI{145}{GHz}$ band picks up the peak frequency of the CMB spectrum. The $\SI{220}{GHz}$ band helps accurate removal of the contamination of thermal emission from the Galactic interstellar dust.
A combined data analysis with the QUIJOTE\cite{rubino2023quijote} experiment is planned. 
QUIJOTE, managed by Instituto de Astrofísica de Canarias (IAC), is a ground-based CMB polarization experiment.
QUIJOTE is operated at the Teide observatory and is located near GroundBIRD.
The entire sky region observed by GroundBIRD is covered by the QUIJOTE experiment. 
The observation bands covered by QUIJOTE are 
11, 13, 17, 19, 30, and 40\, GHz\cite{rubino2023quijote}, which play crucial roles in removing the contamination of the Galactic synchrotron emission from the data. 
Lee et al.(2020)\cite{lee2021forecast} shows that the combined analysis of the data obtained by three years of observations of GroundBIRD with the QUIJOTE data makes it possible to measure the reionization optical depth with an accuracy of $\sigma_{\tau} = 0.012$.
  
This paper reports on the installation of the full set of GroundBIRD detectors, the remote observation system, and the results from on-site performance verification tests and initial observations.
In Section \ref{sec:Instlation}, we describe the telescope; in Section \ref{sec:remote}, we describe the remote observation system; in Section \ref{sec:iniob}, we present the initial observation results; and in Section \ref{sec:conclusion}, we present our conclusions and prospects.  \label{sec:intro}
\section{Detectors, electronics, and measurement methods with MKIDs}\label{sec:Instlation}
\subsection{Detectors}
\begin{figure}[ht]
    \centering
    \begin{subfigure}[t]{0.4\linewidth}
        \centering
        \includegraphics[height=5cm]{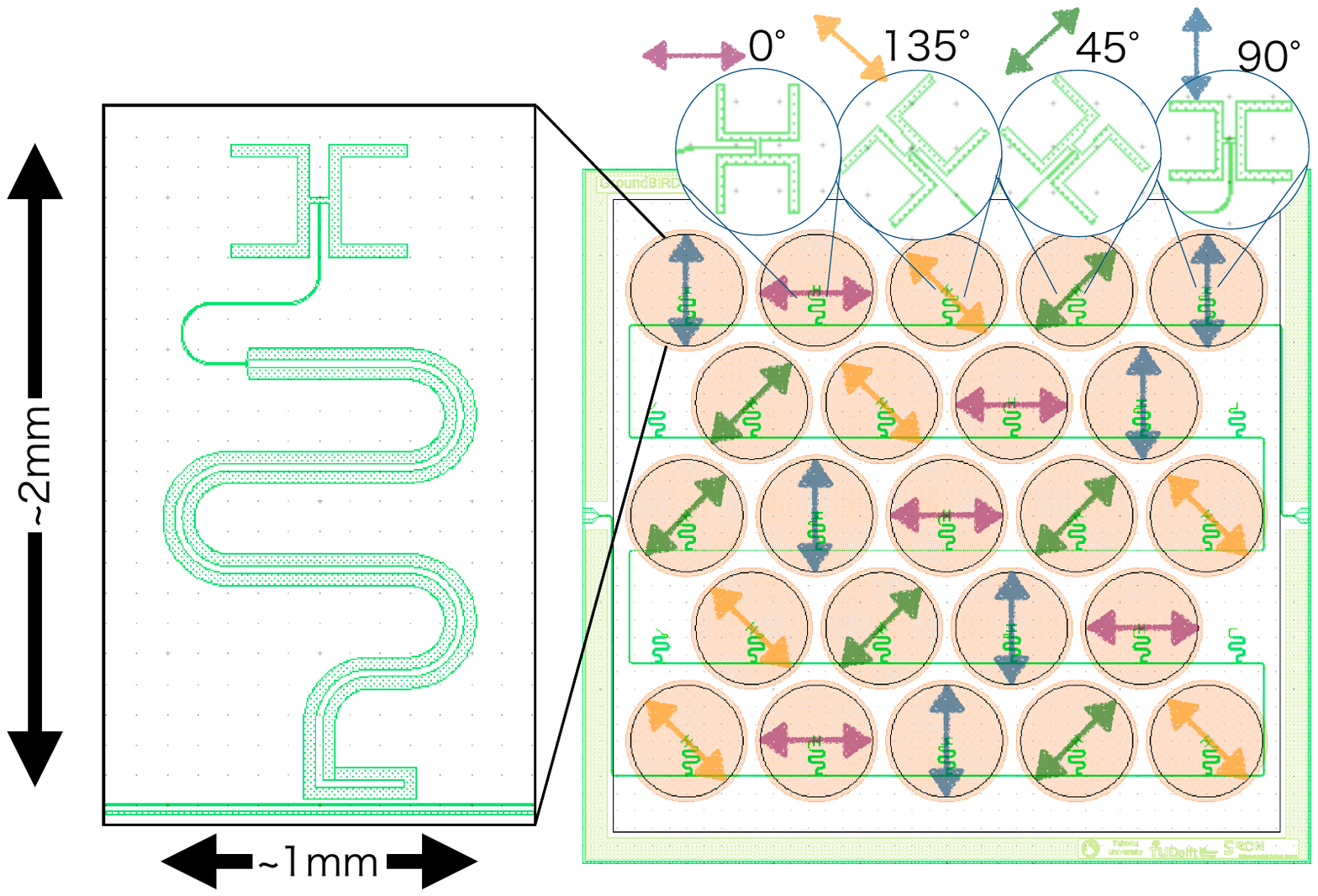}
        \caption{}
        \label{fig:MKIDdesign}
    \end{subfigure}
    \hspace{0.001cm} 
    \begin{subfigure}[t]{0.4\linewidth}
        \centering
        \includegraphics[height=5cm]{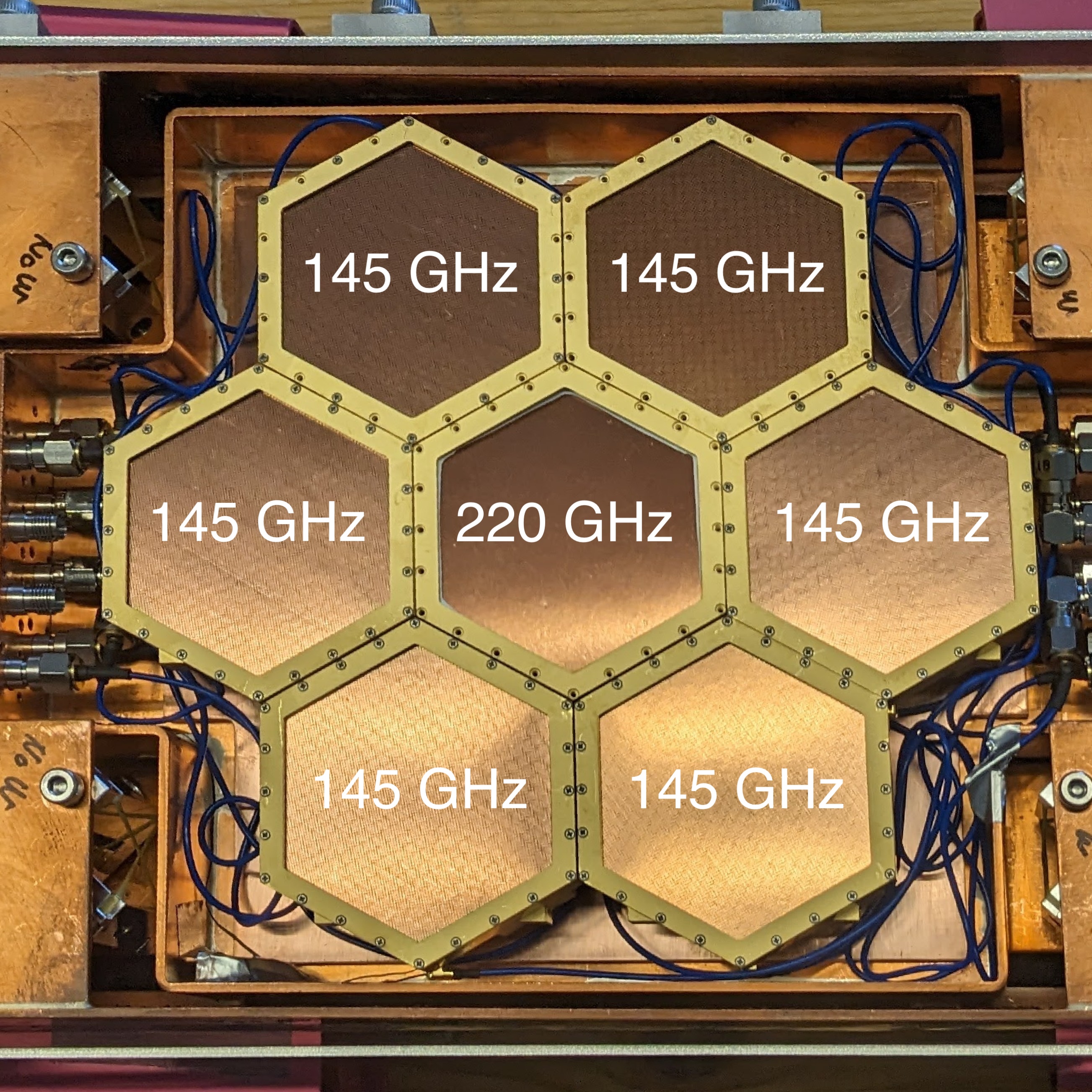}
        \caption{}
        \label{fig:new_focalplane}
    \end{subfigure}
    \caption[example] 
    {\label{fig:MKIDdesign&new_focalplane} 
(a) The design schematic of the MKIDs consists of a hybrid of Al and NbTiN, with detectors equipped with antennas oriented in four directions at 45-degree intervals for measuring Stokes parameters. Each detector measures approximately 2 mm by 1 mm.
(b) Detector arrays installed on the focal plane. The central array is designed for the 220 GHz band, while the surrounding six arrays are designed for the 145 GHz band.}
\end{figure}

   Microwave kinetic inductance detectors (MKIDs) are employed as the focal plane detectors of the GroundBIRD.  
   MKIDs are devices capable of reading out multiple detectors in a single readout line by shifting each resonance frequency. Our focal plane configuration includes seven arrays, each equipped with 23 detectors optimized for polarization observations. The resonance frequencies of these detectors are fine-tuned using the electromagnetic field simulator SONNET, allowing our MKIDs readout system to operate effectively within a 200 MHz bandwidth\cite{ishitsuka2016front} of readout FPGA board.
   To minimize two-level-system noise (TLS noise) contamination of observations, NbTiN/Al hybrid MKIDs are used. 
   The MKID designs are optimized so as to minimize the contamination of the TLS noise and maximize the sensitivity \cite{Kutsuma_mkid_design, kutsuma2021development}(Fig.~\ref{fig:MKIDdesign}).
   The MKID arrays were fabricated at the Netherlands Institute for Space Research (SRON), and were installed in May 2023 (Fig.~\ref{fig:new_focalplane}) following a comprehensive performance verification test conducted in the laboratory. We have confirmed that the time constant of an MKID chip is a few tens of microseconds, ensuring that the response is fast enough for our sampling time. The noise equivalent power at a dark environment (at 800 mK) was approximately $2\times 10^{-17} {\rm W/\sqrt{\rm Hz}}$, which is fit our requirement\cite{Kutsuma_mkid_design}. 
   
   \subsection{Electronics}
       \begin{figure}[ht]
    \begin{center}
    \begin{tabular}{c} 
    \includegraphics[height=4cm]{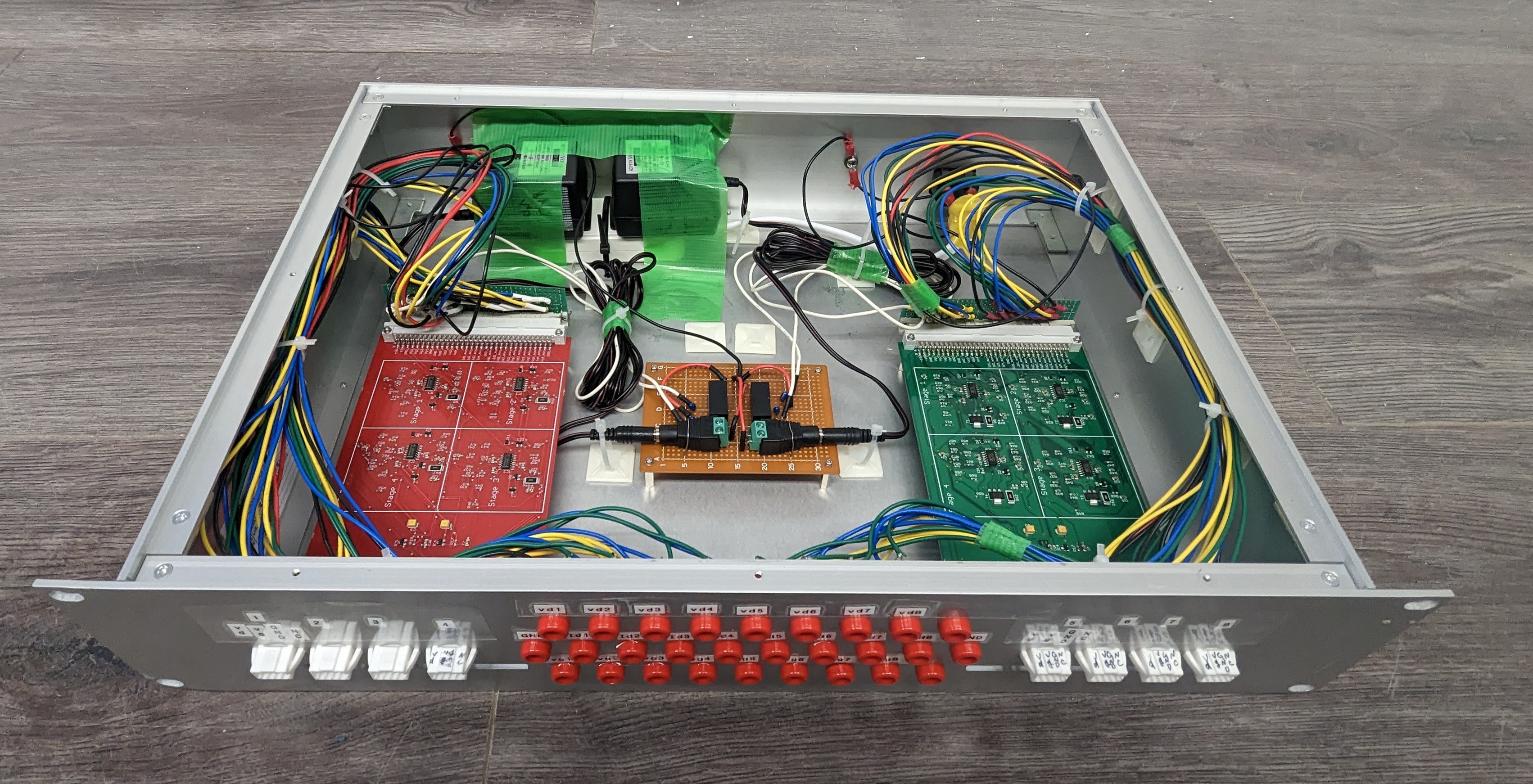}
    \end{tabular}
    \end{center}
    \caption[example] 
       { \label{fig:HEMT_bias_box} 
    A photo of the ultra-low temperature amplifier control system. This system is designed to manage eight Low Noise Amplifiers, of which we utilize seven, leaving one as a spare.}
    \end{figure} 
     An ultra-low temperature amplifier control system (Fig.~\ref{fig:HEMT_bias_box}) was developed to manage the seven Low Noise Amplifiers. This system incorporates the LNF-PS\textunderscore{}EU2\cite{lnf-ps_eu2}, a constant current power supply specifically designed for Low Noise Amplifiers, ensuring stable operation under observational conditions.
        \begin{figure}[ht]
    \centering
    \begin{subfigure}[b]{0.4\linewidth}
        \centering
        \includegraphics[height=5cm]{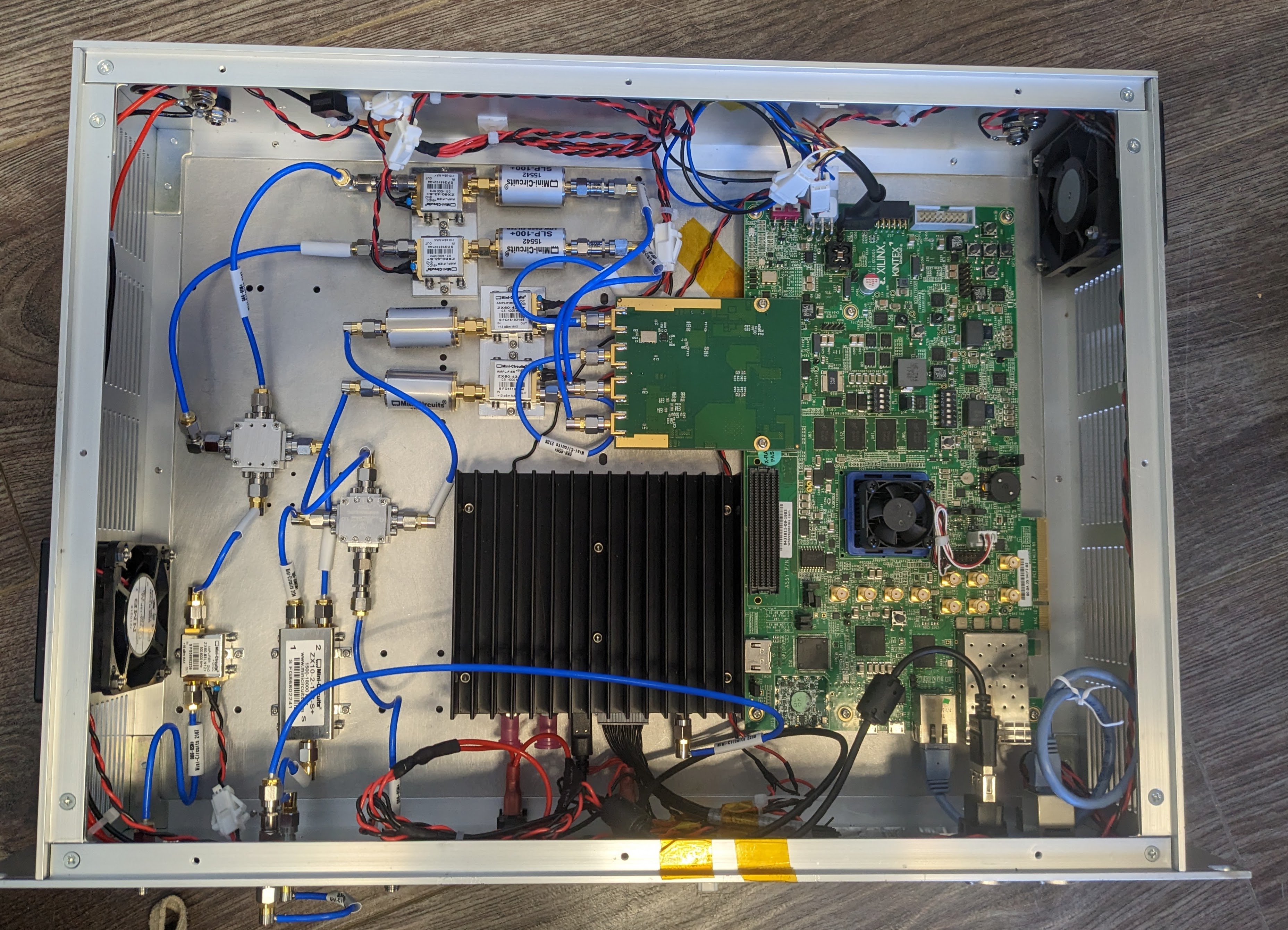}
        \caption{}
        \label{fig:readoutbox}
    \end{subfigure}
    \hspace{0.001cm} 
    \begin{subfigure}[b]{0.4\linewidth}
        \centering
        \includegraphics[height=5cm]{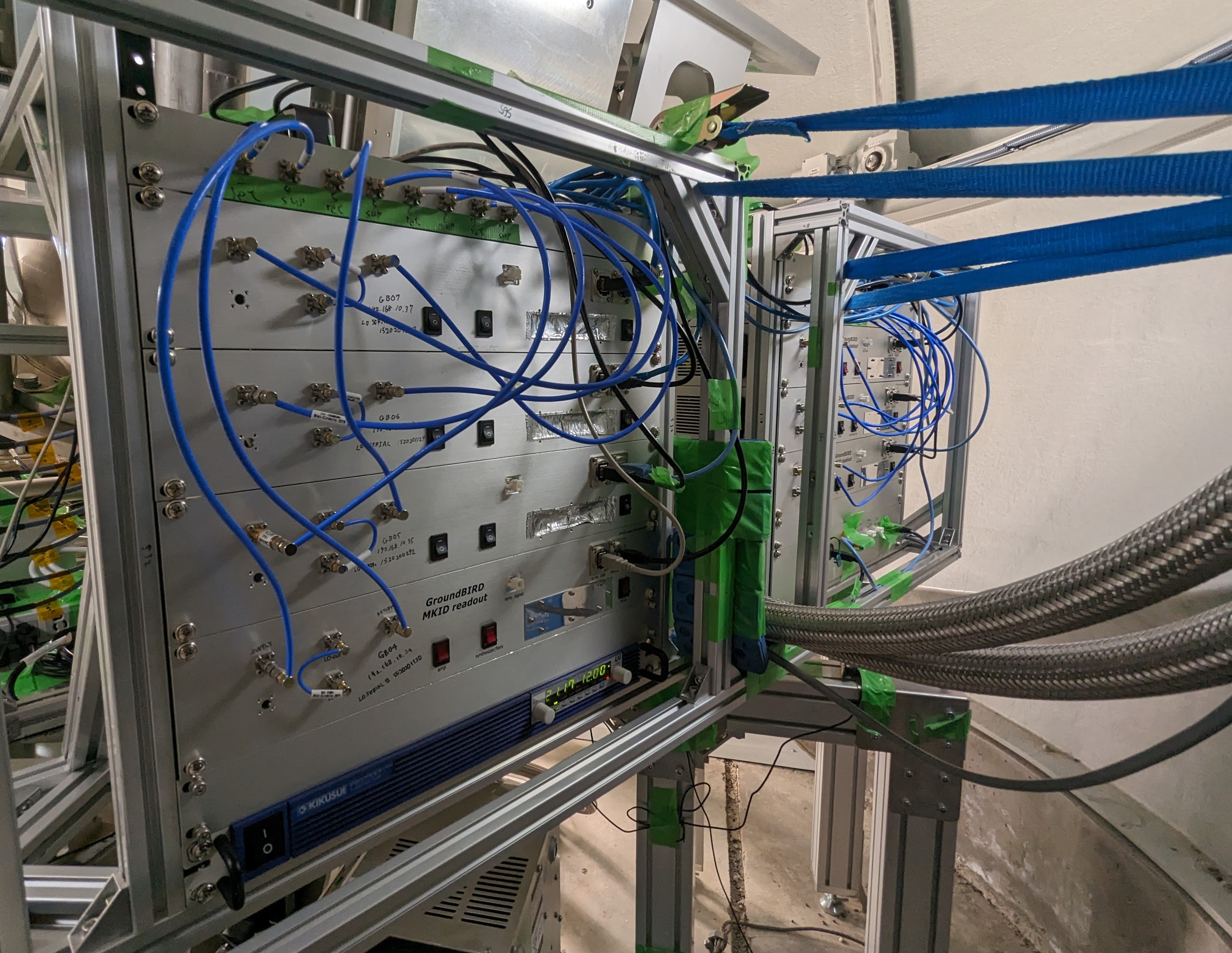}
        \caption{}
        \label{fig:7readoutbox}
    \end{subfigure}
    \caption[example] 
    { \label{fig:readoutbox&7readoutbox} 
(a) A photo of GroundBIRD readout box. (b) A photo of all readout boxes installed.}
    \end{figure}

    A readout system is composed of seven data acquisition (DAQ) readout boxes for recording signals from each MKID sent from seven arrays (Fig.~\ref{fig:readoutbox&7readoutbox}) and two DAQ readout boxes for recording azimuth and elevation angles of the pointing direction. 
    In order to synchronize the timing in all DAQ readout boxes, the azimuth DAQ issues a pulse signal named ``sync-pulse''. This pulse is transmitted to DAQ boxes. Using the pulse information, all the time stream data are aligned. The azimuth and elevation angles are interpolated to obtain the angle values at each MKID clock timing\cite{honda2020site}. 


\subsection{Measurement methods with MKIDs}
    \begin{figure}[b]
    \begin{center}
    \begin{tabular}{c} 
    \includegraphics[height=3.5cm]{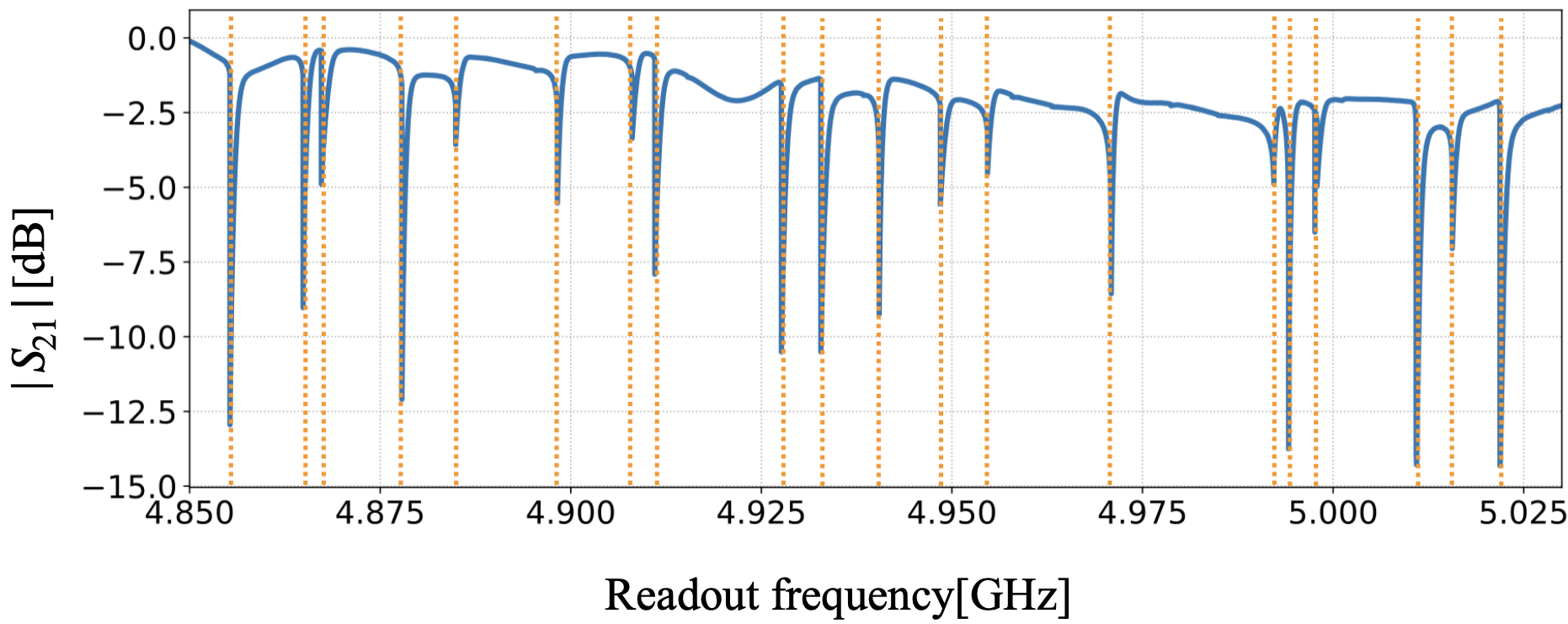}
    \end{tabular}
    \end{center}
    \caption[example] 
       { \label{fig:sweep}
    Multiple resonances of a single array. The horizontal axis represents the readout frequency, and the vertical axis represents the readout power. The solid blue line shows the readout power, while the dotted orange lines indicate the resonance frequencies of each element.}
    \end{figure} 
 Figure~\ref{fig:sweep} shows an example of resonance frequency distribution for one MKID array within a readout frequency band of 200 MHz.
 The vertical axis represents the absolute value of $S_{21}$\cite{gao2008physics}, while the horizontal axis represents the frequency of the readout power.
 In this example, twenty resonances are identified within the readout frequency band. By sweeping the frequency of the readout power, the resonance frequency of each detector pixel is identified. 
 For each pixel, the amplitude of the readout powers are 
 optimized to maximize the amplitude of the resonance feature without altering its shape of the resonance feature (see the left panel of Fig.~\ref{fig:slack}). 
 During observations, the optimized power at the resonance frequency for each detector pixel is simultaneously provided for all detector pixels mounted on the same MKID array. 
 The amplitude and phases of $S_{21}$ for each detector pixel are measured at a 1 kHz sampling rate. 
 The resonant frequencies of MKIDs undergo gradual shifts due to factors such as fluctuations in the amplitude of the atmospheric emission, which acts as the main optical heat load for detectors. Therefore, the resonance frequency of each detector is calibrated every hour by measuring the resonance shape around the originally defined resonance frequency. 
 In subsequent observations, 
 the readout powers are given at the recalibrated resonance frequency. 
 The calibration of the resonance frequency is performed every hour, and both the calibration and the measurement of $S_{21}$ at the calibrated resonance frequency are automated.



\section{Remote observation systems}
All tasks, including dome and telescope operations and data acquisition (DAQ), are managed remotely from locations in Japan, Spain, South Korea, and the United Kingdom.
We have developed a remote observation system for the long-term monitoring of the observations with the GroundBIRD telescope.  
This system enables monitoring of various conditions, including atmospheric conditions in the GroundBIRD observation area, weather at the observatory, internal conditions of the GroundBIRD telescope such as temperature and pressure, as well as conducting visual and audio checks on the telescope.

\begin{figure}[hb]
    \centering
    \begin{subfigure}[b]{0.4\linewidth}
        \centering
        \includegraphics[height=5cm]{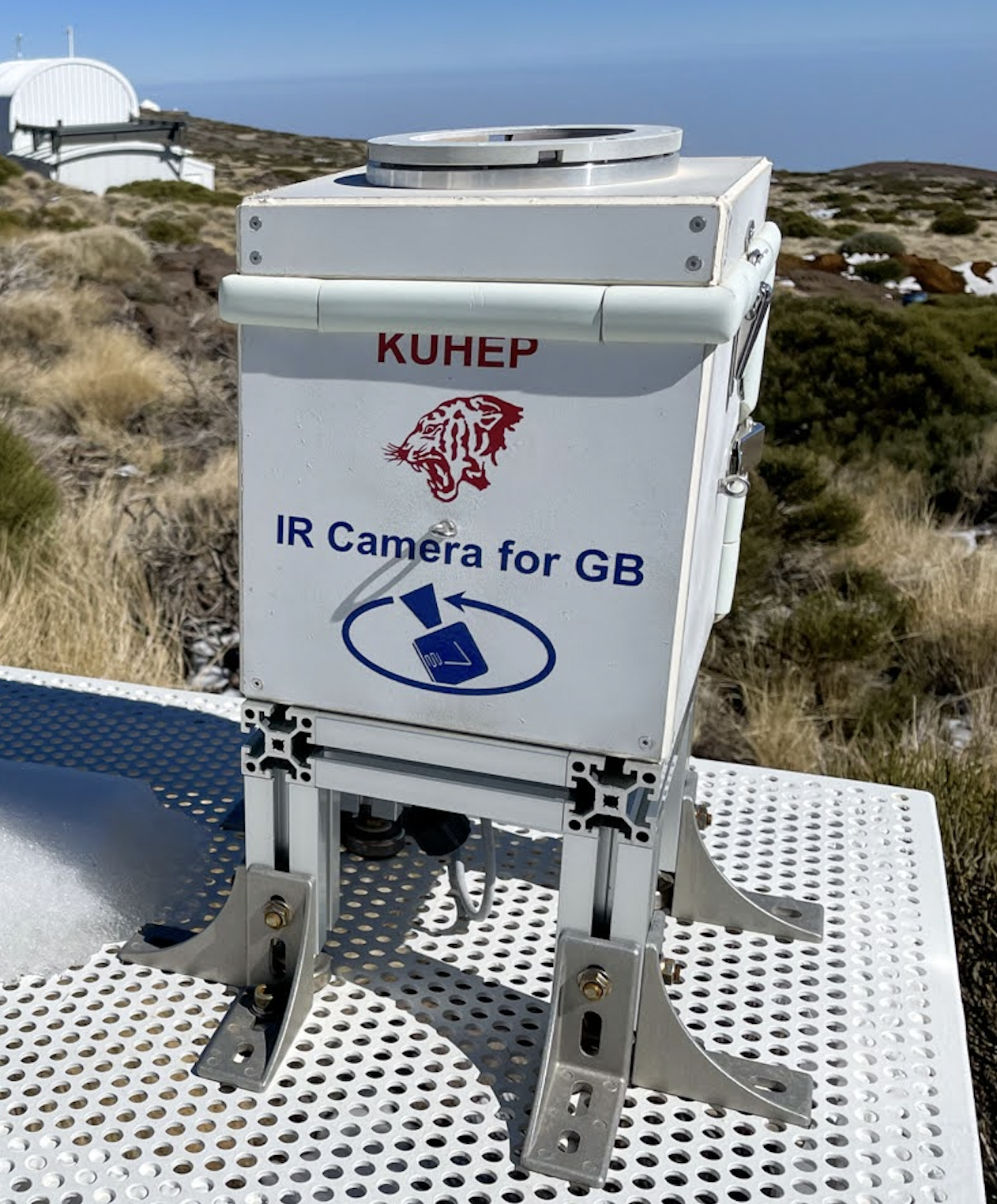}
        \caption{}
        \label{fig:GBIRcamera}
    \end{subfigure}
    \hspace{0.001cm} 
    \begin{subfigure}[b]{0.4\linewidth}
        \centering
        \includegraphics[height=6cm]{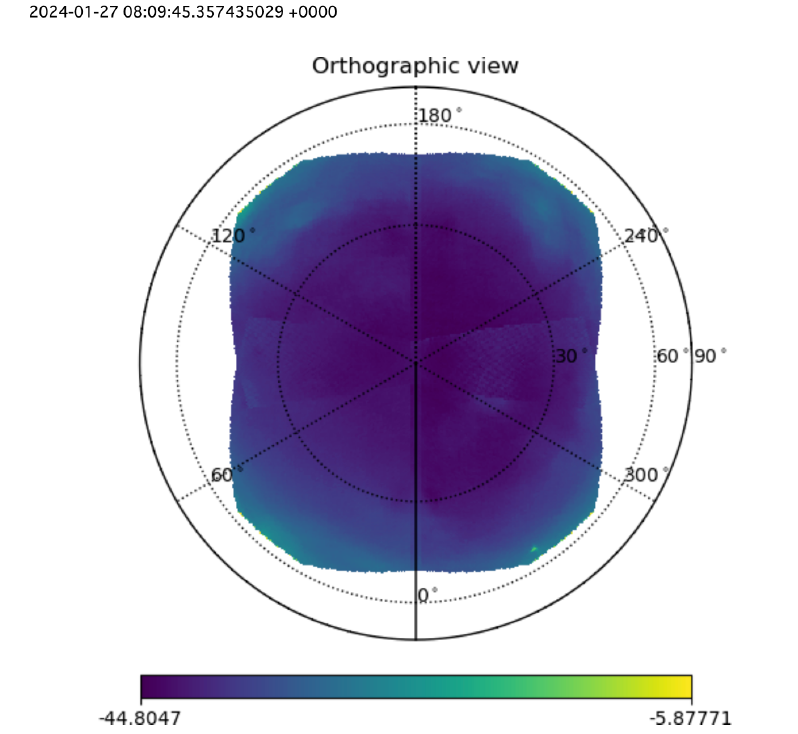}
        \caption{}
        \label{fig:IRcamera_monitor}
    \end{subfigure}
    \caption[example] 
    {\label{fig:GBIRcamera&IRcamera_monitor}
    (a) An exterior view of the GroundBIRD infrared camera, devised by the GroundBIRD team, which functions as an infrared monitoring system for detecting clouds at the Teide Observatory. (b) Web-monitor system, enabling convenient online monitoring of the GroundBIRD infrared camera data.}
    \end{figure}

The GroundBIRD infrared camera, devised by the GroundBIRD team, functions as an infrared monitoring system for detecting clouds at the Teide Observatory\cite{Hoyong_IRcam}. It is built around the FLIR Lepton 3.5\cite{Lepton} and PureThermal 2 modules\cite{PureThermal}, providing a crucial tool for cloud surveillance (Fig.~\ref{fig:GBIRcamera}). Additionally, this system features web-based accessibility, allowing for convenient monitoring online (Fig.~\ref{fig:IRcamera_monitor}).

\begin{figure} [ht]
   \begin{center}
   \begin{tabular}{c} 
   \includegraphics[height=5cm]{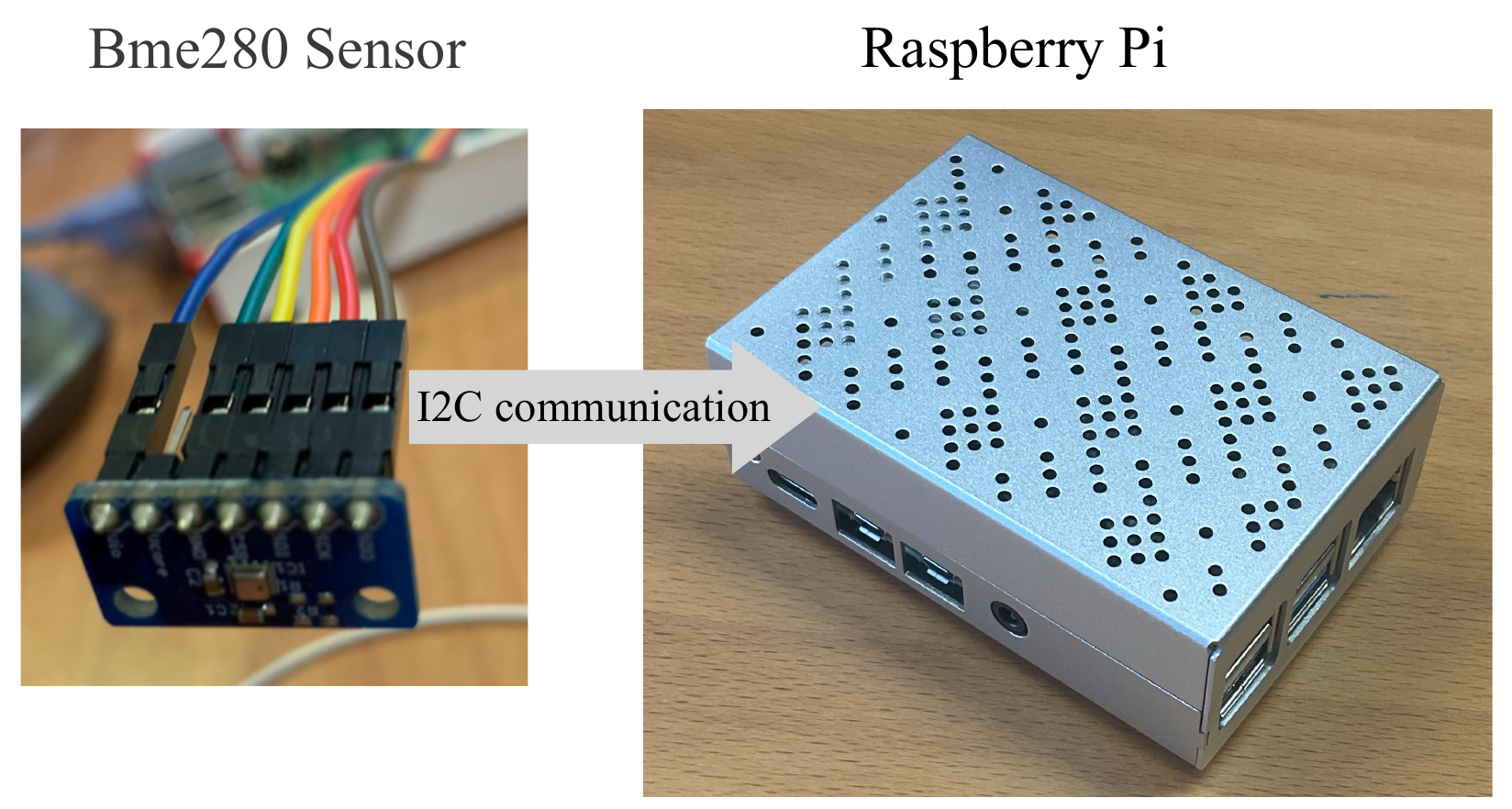}
   \end{tabular}
   \end{center}
   \caption[example] 
   {\label{fig:bme280} 
The BME280 sensor and Raspberry Pi system are used for monitoring environmental conditions inside and around the GroundBIRD dome. The BME280 sensor measures temperature, humidity, and atmospheric pressure, while the Raspberry Pi facilitates data collection and communication. This system helps ensure safe long-term observations by issuing alerts and automatically controlling the telescope dome.}
   \end{figure}
BME280 sensors have been installed to measure temperature, humidity, and atmospheric pressure both inside and around the GroundBIRD dome. 
Data collection is facilitated using Raspberry Pi devices, which communicate with sensors via I2C communication (Fig.~\ref{fig:bme280}).
The Raspberry Pi devices perform dual functions: collecting data and issuing alerts. If humidity exceeds a predefined threshold, an alert is triggered, and a command is automatically sent to the dome server to close the telescope dome. These systems are instrumental in ensuring the safety of long-term observations.

To ensure the safety of observations, it is crucial to check multiple sources of information on the observatory. Grafana\cite{grafana}, an open-source visualization tool, plays a key role in this process. Our records encompass a range of telescope-related data, including meteorological factors such as wind velocity, humidity, and precipitable water vapor (PWV), as well as cryostat parameters like temperature and vacuum levels. By visualizing these records through Grafana, we can make informed decisions to maintain observation safety.
   \begin{figure} [hpb]
   \begin{center}
   \begin{tabular}{c} 
   \includegraphics[height=3.3cm]{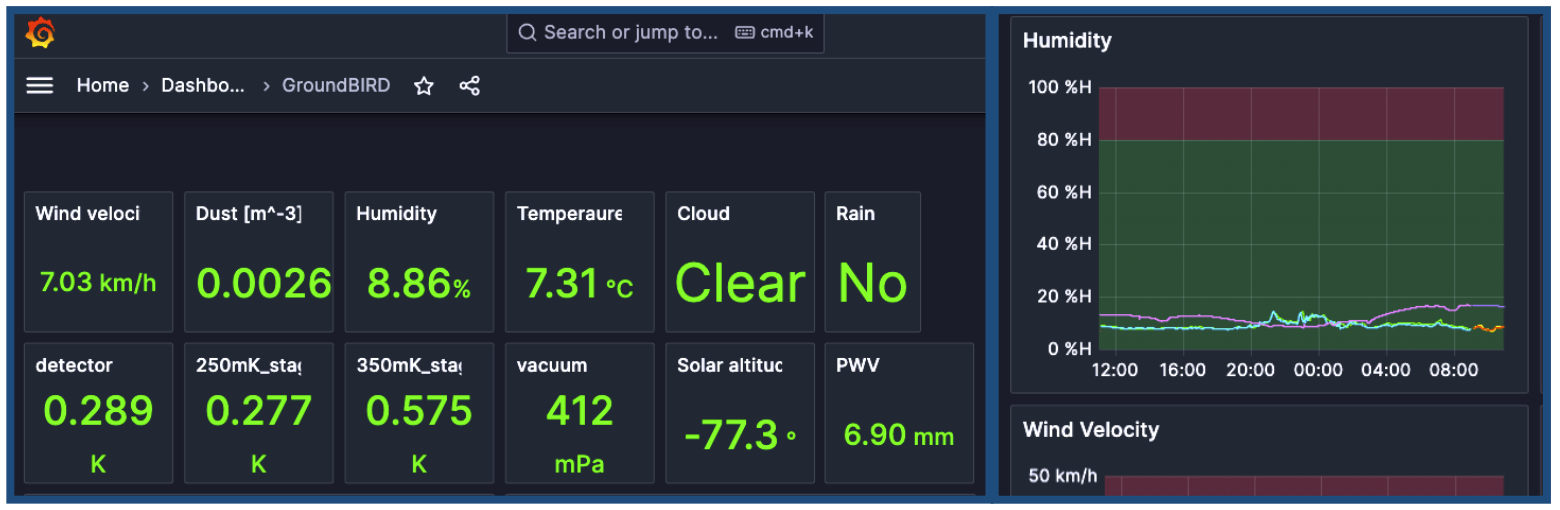}
   \end{tabular}
   \end{center}
   \caption[example]
   {\label{fig:grafana}
Visualization of telescope environmental data using Grafana. The monitor displays meteorological information such as wind speed and humidity, alongside telescope parameters such as detector temperature and vacuum levels.}
   \end{figure}

\begin{figure}[hb]
\begin{center}
\begin{tabular}{c} 
\includegraphics[height=5cm]{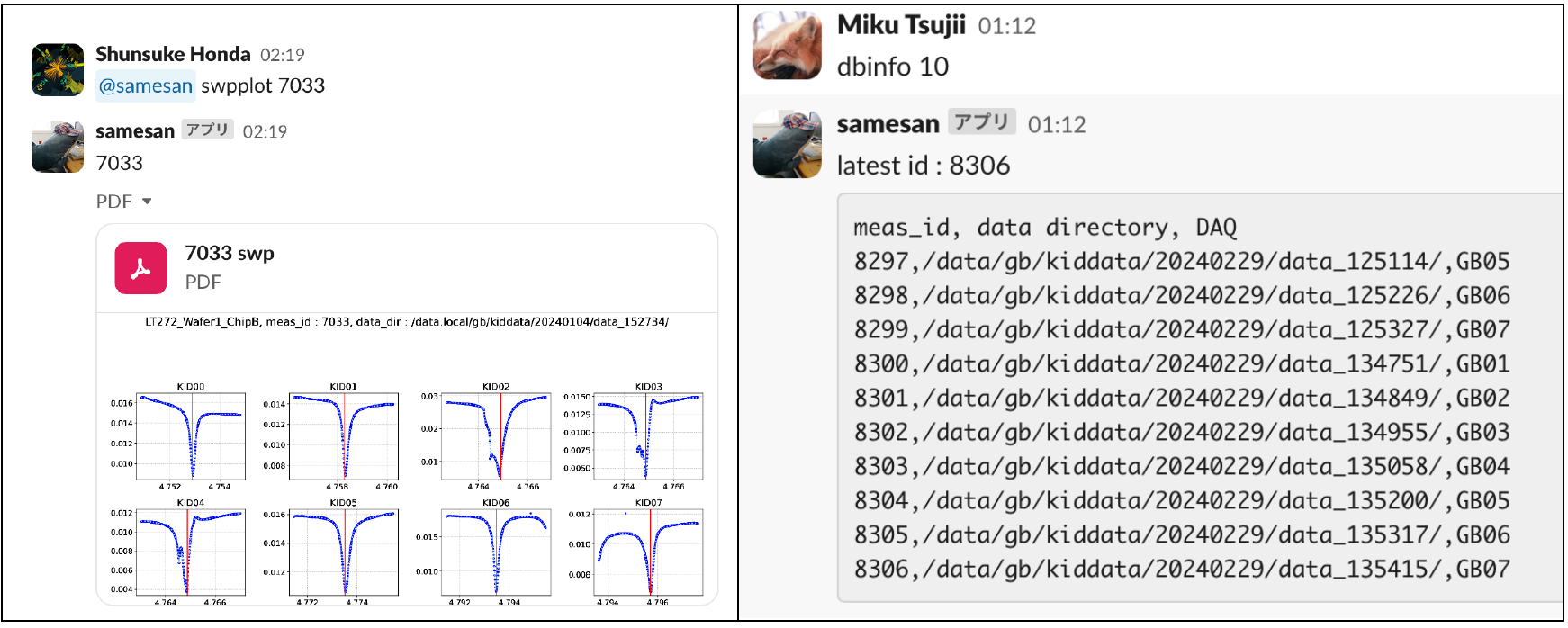}
\end{tabular}
\end{center}
\caption[example] 
   { \label{fig:slack}
Examples of using Slack: The process of checking the MKID sweeps is shown here. Additionally, Slack can be used to access telescope visuals, observatory information, and more.}
\end{figure}
We have developed a telescope monitoring system using the team communication platform Slack\cite{slack}. This system is built using Slackbot\cite{slackbot} and Slacker\cite{slacker}, providing straightforward access to visual information from the telescope, including telescope recordings and coordinates of targeted celestial bodies employed for calibration.
 We check for issues in the automated process through Slack (Fig.~\ref{fig:slack}).\label{sec:remote}
\section{on-site performance verification tests} 
\begin{figure}[H]
\begin{center}
\begin{tabular}{c} 
\includegraphics[height=7.cm]{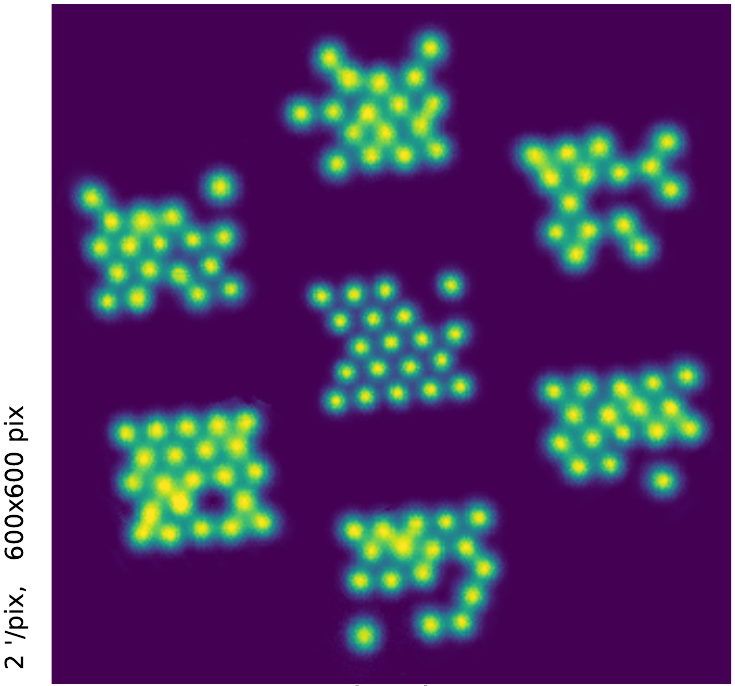}
\end{tabular}
\end{center}
\caption[example] 
   { \label{fig:moon}
 A map centered on the beam, constructed using nearly 80\% of the detector pixels. Each detector's response has been normalized.}
\end{figure}
We performed mapping observations of the moon
as the first step of the on-site performance verification tests. 
Figure~\ref{fig:moon} shows the intensity distributions of the moon measured by each MKID pixel.
The intensities shown in Fig.~\ref{fig:moon} are normalized by the peak intensities for each MKID pixel.
It shows that the intensity distribution of the moon was successfully reconstructed by all available detector pixels.
We can also see that the intensity maps of the moon obtained by 145 GHz band detectors are larger than those obtained by 220 GHz band detectors, as expected.
This reflects that the beam width of the 145 GHz detectors is comparable to the apparent size of the moon, whereas the beam width of 220 GHz detectors is smaller than the moon size\cite{jo2023simulation}.    
Details of the pointing calibrations based on the moon observations are provided in Sueno et al. (2024)\cite{sueno2024pointing}. 
Through these observations, we successfully correlated the resonance frequencies with the positions of the detectors (Fig.~\ref{fig:moon}).

\begin{figure}[b]
 \centering
 \begin{subfigure}[hb]{0.4\linewidth}
 \centering
     \includegraphics[height=3.5cm]{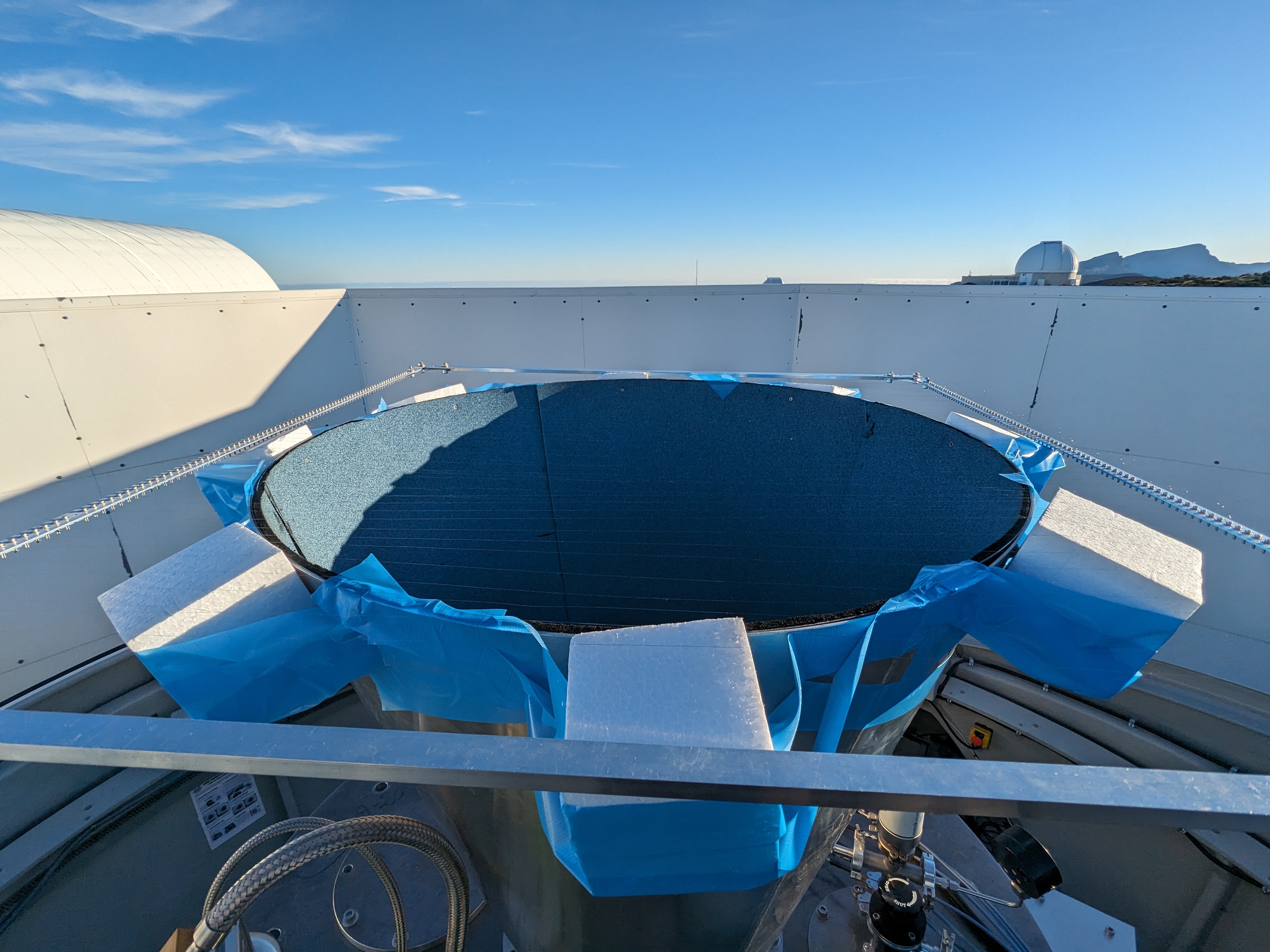}
     \caption{}
     \label{fig:wire_obs}
 \end{subfigure}
 \hspace{0.001cm} 
 \begin{subfigure}[hb]{0.5\linewidth}
 \centering
     \includegraphics[height=4cm]{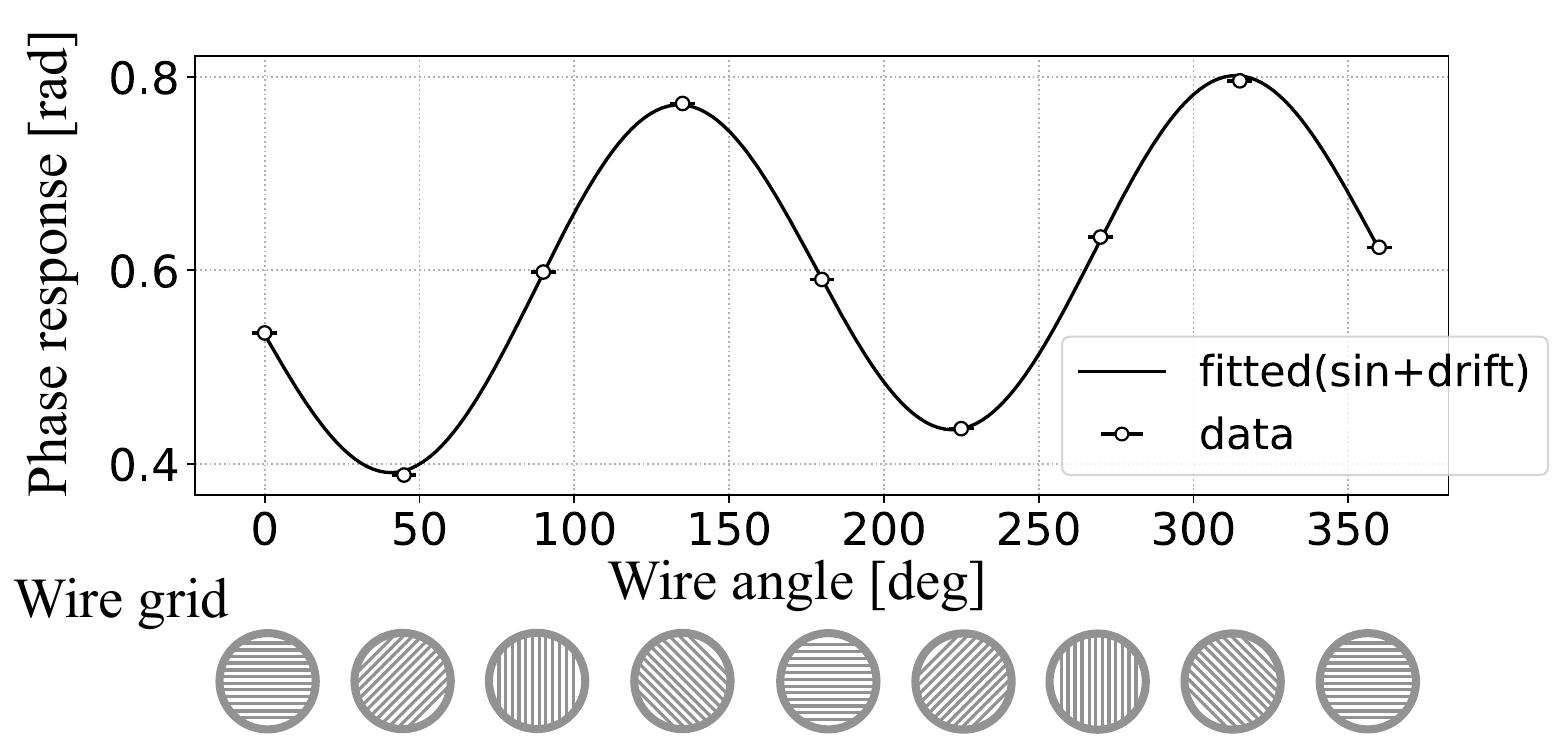}
     \caption{}
     \label{fig:res}
 \end{subfigure}
 \caption[example] 
 {\label{fig:wire_obs&/res}
 (a) Sparse wire grid placed on the baffle, using foam polystyrene as stands. (b) Response of an MKID when the wire grid is rotated. The signal modulated by the wires can be observed.}
 \end{figure}
\begin{figure}[t]
    \begin{center}
    \begin{tabular}{c} 
    \includegraphics[height=4.8cm]{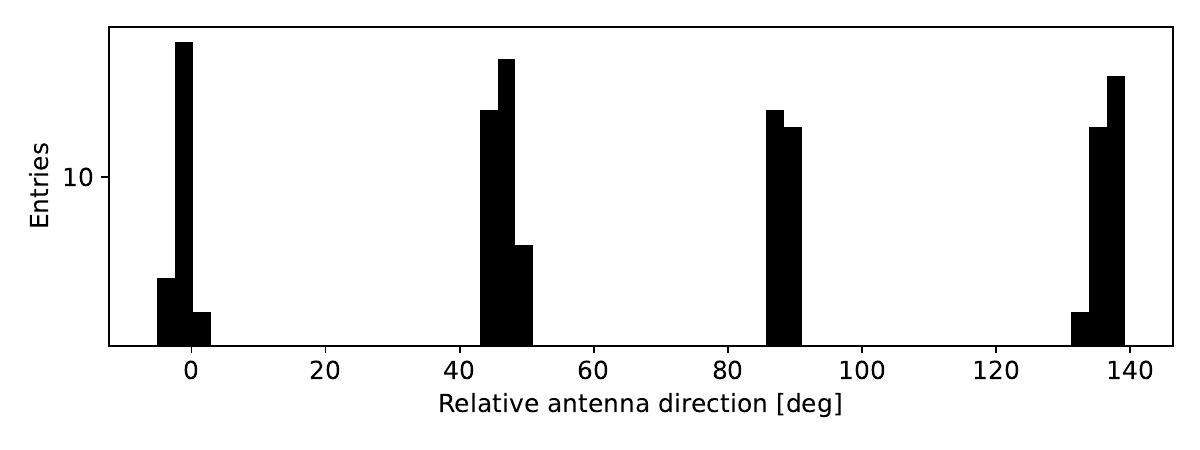}
    \end{tabular}
    \end{center}
    \caption[example] 
       { \label{fig:hist}
    Analysis of the polarization response angles of the detector. The results are divided as designed at $0^{\circ}, 45^{\circ}, 90^{\circ},$ and $135^{\circ}$, using one detector as the reference.}
    \end{figure}

Quick checks of the sensitivity of each detector pixel were performed using a sparse wire grid.
The sparse wire grid was constructed by stretching 50 tungsten wires, each with a diameter of 0.1 mm and spaced 16 mm apart, on a square aluminum frame approximately 1 m in diameter.
The wire grid was set on top of the baffle in the state that the GroundBIRD telescope pointed at the zenith. 
To install the sparse wire grid, we attached foam polystyrene to the GroundBIRD baffle. The foam polystyrene served as guides for positioning and as stands for the wire grid. Using the foam polystyrene as a guide, the grid was manually positioned, achieving an angular accuracy of approximately $\pm 4^{\circ}$ (Fig.~\ref{fig:wire_obs}).
Data were acquired for each setting of the wire grid. 
By rotating the wire grid around the optical axis in $45^{\circ}$ intervals, the measurements were repeated for one complete round.
An example of the obtained data is shown in Fig.~\ref{fig:res}. 
The data for each detector were fitted with trigonometric and cubic functions (Fig.~\ref{fig:res}). The trigonometric function corresponds to the signal from the wires, while the cubic function accounts for baseline noise. 
We confirmed that the detectors are sensitive to linearly polarized signals.
From the fitting results, the direction of the polarization signals to which each detector pixel is sensitive was extracted. 
Figure~\ref{fig:hist} shows the histogram of detector pixels as a function of the direction of the sensitive polarization signal.
Although the accuracies of the measurements were limited, Fig.~\ref{fig:hist} shows that each detector is sensitive to the expected direction of the polarized signal.

    

\begin{figure}[b]
\begin{center}
\begin{tabular}{c} 
\includegraphics[height=6.cm]{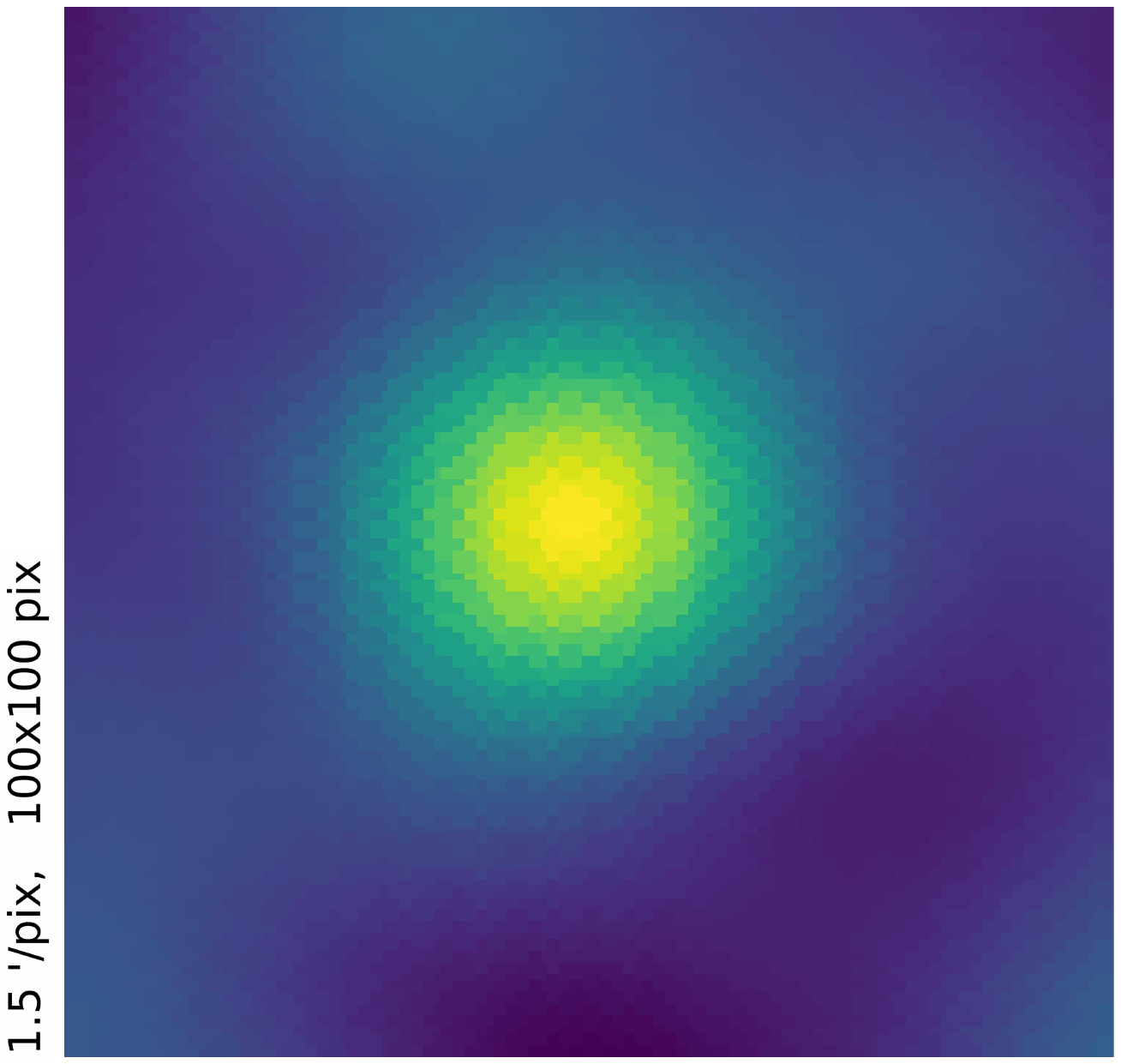}
\end{tabular}
\end{center}
\caption[example] 
   { \label{fig:Jupiter}
An image of Jupiter captured by measuring the phase response of a single detector pixel in the 145 GHz band. The map, centered on Jupiter, displays an area of $100 \times 100$ pixels, each with a width of $\ang{;1.5;}$. The image has been smoothed using the HEALPix package\cite{gorski2005healpix} with a FWHM of $\ang{0.6}$\cite{lee2021forecast}.}
\end{figure}
Furthermore, we have successfully obtained the intensity maps of Jupiter. 
Figure~\ref{fig:Jupiter} shows a map obtained by measuring the phase response of a single detector pixel as Jupiter passed through the field of view.

\label{sec:iniob}
\section{conclusion and prospects}
GroundBIRD is a ground-based CMB experiment designed to observe the polarization patterns imprinted on large angular scales.
We have successfully initiated the GroundBIRD science observation run using the full MKID arrays.
These observations are conducted remotely from multiple countries, supported by the newly developed remote observation system. 
Initial validations, including polarization response tests and mappings of the moon and Jupiter, have been completed successfully. 
Concurrently with data acquisition progress,
we are working on upgrading the instrument to improve measurement accuracy to achieve the scientific aim of the GroundBIRD.\label{sec:conclusion}

\acknowledgments 
This work was supported by MEXT KAKENHI Grant Number JP18H05539 and  JSPS KAKENHI Grant Numbers JP15H05743, JP20K20927, JP20KK0065, JP21H04485, JP21K03585, JP22H04913, and JP24H00224, JSPS Bilateral Program Numbers JPJSBP120219943 and JPJSBP120239919, and also supported by JSPS Core-to-Core Program JPJSCCA20200003.
M.T. also acknowledges support from Graduate Program on Physics for the Universe (GP-PU), Tohoku University, and JST SPRING, Grant Number JPMJSP2114.
Y.S. also acknowledges support from Grant Number JP21J20290.
T.T. also acknowledges support from JST SPRING, Grant Number JPMJSP2114.
K.L. and E.W. also acknowledge support from Grant Number 2022R1A2B5B02001535 by the National Research Foundation of Korea (NRF).
We thank Victor González Escalera, \'{A}ngeles P\'{e}rez de Taoro (the Instituto de Astrof\'{i}sica de Canarias), and the staff of Teide Observatory for supporting the maintenance and operation of GroundBIRD.
We also express special appreciation to Sander van Loon (SRON) for Lens array assembly and Akira Endo (Delft University of Technology) for providing MKIDs fabricated.
\label{sec:intro}  
\bibliography{report} 

\begin{thebibliography}{10}

\bibitem{1965PenziasWilsonCMB}
Penzias, A.~A. and Wilson, R.~W., ``A measurement of excess antenna temperature at 4080 mc/s.,'' {\em apj}~{\bf 142},  419--421 (jul 1965).

\bibitem{dicke1965cosmic}
Dicke, R.~H., Peebles, P. J.~E., Roll, P.~G., and Wilkinson, D.~T., ``Cosmic black-body radiation.,'' {\em Astrophysical Journal}~{\bf 142},  414--419 (1965).

\bibitem{kamionkowski1997probe}
Kamionkowski, M., Kosowsky, A., and Stebbins, A., ``A probe of primordial gravity waves and vorticity,'' {\em Physical Review Letters}~{\bf 78}(11),  2058 (1997).

\bibitem{zaldarriaga1997all}
Zaldarriaga, M. and Seljak, U., ``All-sky analysis of polarization in the microwave background,'' {\em Physical Review D}~{\bf 55}(4),  1830 (1997).

\bibitem{starobinskii1979spectrum}
Starobinskii, A., ``Spectrum of relict gravitational radiation and the early state of the universe,'' {\em JETP Letters}~{\bf 30}(11),  682--685 (1979).

\bibitem{starobinsky1980new}
Starobinsky, A.~A., ``A new type of isotropic cosmological models without singularity,'' {\em Physics Letters B}~{\bf 91}(1),  99--102 (1980).

\bibitem{sato1981first}
Sato, K., ``First-order phase transition of a vacuum and the expansion of the universe,'' {\em Monthly Notices of the Royal Astronomical Society}~{\bf 195}(3),  467--479 (1981).

\bibitem{guth1982fluctuations}
Guth, A.~H. and Pi, S.-Y., ``Fluctuations in the new inflationary universe,'' {\em Physical Review Letters}~{\bf 49}(15),  1110 (1982).

\bibitem{allison2015towards}
Allison, R., Caucal, P., Calabrese, E., Dunkley, J., and Louis, T., ``Towards a cosmological neutrino mass detection,'' {\em Physical Review D}~{\bf 92}(12),  123535 (2015).

\bibitem{lewis2000efficient}
Lewis, A., Challinor, A., and Lasenby, A., ``Efficient computation of cosmic microwave background anisotropies in closed friedmann-robertson-walker models,'' {\em The Astrophysical Journal}~{\bf 538}(2),  473 (2000).

\bibitem{collaboration2020planck}
Collaboration, P., Aghanim, N., Akrami, Y., Ashdown, M., Aumont, J., Baccigalupi, C., Ballardini, M., Banday, A., Barreiro, R., Bartolo, N., et~al., ``Planck 2018 results. vi. cosmological parameters,'' (2020).

\bibitem{bennett2013nine}
Bennett, C.~L., Larson, D., Weiland, J.~L., Jarosik, N., Hinshaw, G., Odegard, N., Smith, K., Hill, R., Gold, B., Halpern, M., et~al., ``Nine-year wilkinson microwave anisotropy probe (wmap) observations: final maps and results,'' {\em The Astrophysical Journal Supplement Series}~{\bf 208}(2),  20 (2013).

\bibitem{rubino2023quijote}
Rubi{\~n}o-Mart{\'\i}n, J., Guidi, F., G{\'e}nova-Santos, R., Harper, S., Herranz, D., Hoyland, R., Lasenby, A., Poidevin, F., Rebolo, R., Ruiz-Granados, B., et~al., ``Quijote scientific results--iv. a northern sky survey in intensity and polarization at 10--20 ghz with the multifrequency instrument,'' {\em Monthly Notices of the Royal Astronomical Society}~{\bf 519}(3),  3383--3431 (2023).

\bibitem{lee2021forecast}
Lee, K., G{\'e}nova-Santos, R.~T., Hazumi, M., Honda, S., Kutsuma, H., Oguri, S., Otani, C., Peel, M.~W., Sueno, Y., Suzuki, J., et~al., ``A forecast of the sensitivity on the measurement of the optical depth to reionization with the groundbird experiment,'' {\em The Astrophysical Journal}~{\bf 915}(2),  88 (2021).

\bibitem{ishitsuka2016front}
Ishitsuka, H., Ikeno, M., Oguri, S., Tajima, O., Tomita, N., and Uchida, T., ``Front--end electronics for the array readout of a microwave kinetic inductance detector towards observation of cosmic microwave background polarization,'' {\em Journal of Low Temperature Physics}~{\bf 184},  424--430 (2016).

\bibitem{Kutsuma_mkid_design}
Kutsuma, H. et~al. In preparation.

\bibitem{kutsuma2021development}
Kutsuma, H., {\em Development of novel calibration methods and performance forecaster of cutting-edge superconducting detector MKIDs for CMB experiments}, PhD thesis, Tohoku University (2021).

\bibitem{lnf-ps_eu2}
``Lnf-ps\_eu2.'' \url{https://lownoisefactory.com/product/lnf-ps_eu2/}.

\bibitem{honda2020site}
Honda, S., Choi, J., G{\'e}nova-Santos, R.~T., Hattori, M., Hazumi, M., Ikemitsu, T., Ishida, H., Ishitsuka, H., Jo, Y., Karatsu, K., et~al., ``On-site performance of groundbird, a cmb polarization telescope for large angular scale observations,'' in [{\em Ground-based and Airborne Telescopes VIII}{\nolinebreak\hspace{0.1em}]},   {\bf 11445},  1379--1386, SPIE (2020).

\bibitem{gao2008physics}
Gao, J.,  [{\em The physics of superconducting microwave resonators}{\nolinebreak\hspace{0.1em}]}, California Institute of Technology (2008).

\bibitem{Hoyong_IRcam}
Hoyong, J., Yongheon, A., Eunil, W., and Kyungmin, L., ``Development of cloud monitoring system for cosmic microwave background observations,'' in [{\em Journal of the Korean Physical Society}{\nolinebreak\hspace{0.1em}]},   {\bf 80},  88--93 (2022).

\bibitem{Lepton}
``Lepton 3.5.'' \url{https://www.flir.jp/products/lepton/}.

\bibitem{PureThermal}
``Purethermal 2.'' \url{https://groupgets.com/manufacturers/getlab/products/purethermal-2-flir-lepton-smart-i-o-module}.

\bibitem{grafana}
``Grafana.'' \url{https://grafana.com}.

\bibitem{slack}
``Slack.'' \url{https://slack.com/}.

\bibitem{slackbot}
``slackbot.'' \url{https://github.com/lins05/slackbot}.
\newblock Accessed: 6 May 2024.

\bibitem{slacker}
``slacker.'' \url{https://github.com/os/slacker}.
\newblock Accessed: 6 May 2024.

\bibitem{jo2023simulation}
Jo, Y., Choi, J., Hattori, M., Honda, S., Tanaka, T., Tsuji, M., Won, E., and Lee, K., ``Simulation of the optical system in the groundbird telescope,'' {\em Applied optics}~{\bf 62}(20),  5369--5378 (2023).

\bibitem{sueno2024pointing}
Sueno, Y., Baselmans, J., Coppens, A., G{\'e}nova-Santos, R., Hattori, M., Honda, S., Karatsu, K., Kutsuma, H., Lee, K., Nagasaki, T., et~al., ``Pointing calibration of groundbird telescope using moon observation data,'' {\em Progress of Theoretical and Experimental Physics}~{\bf 2024}(2),  023F01 (2024).

\bibitem{gorski2005healpix}
Gorski, K.~M., Hivon, E., Banday, A.~J., Wandelt, B.~D., Hansen, F.~K., Reinecke, M., and Bartelmann, M., ``Healpix: A framework for high-resolution discretization and fast analysis of data distributed on the sphere,'' {\em The Astrophysical Journal}~{\bf 622}(2),  759 (2005).

\end{thebibliography}
\bibliographystyle{spiebib} 

\end{document}